\documentclass[10pt,twocolumn,preprintnumbers,showpacs,superscriptaddress]{revtex4-1}
\usepackage{graphicx,dcolumn,bm,color,amsmath,amssymb,amsfonts,esint}
\usepackage{times}

\begin{document}

\title{Controllable high-fidelity quantum state transfer and entanglement generation in circuit QED}

\author{Peng Xu}
\affiliation{National Laboratory of Solid State Microstructures, School of Physics,
Nanjing University, Nanjing 230039, China}

\author{Xu-Chen Yang}
\affiliation{Guangdong Provincial Key Laboratory of Quantum Engineering and Quantum
Materials, and School of Physics\\ and Telecommunication Engineering, South
China Normal University, Guangzhou 510006, China}

\author{Feng Mei}
\affiliation{National Laboratory of Solid State Microstructures, School of Physics,
Nanjing University, Nanjing 230039, China}

\author{Zheng-Yuan Xue}
\affiliation{Guangdong Provincial Key Laboratory of Quantum
Engineering and Quantum Materials, and School of Physics\\ and
Telecommunication Engineering, South China Normal University,
Guangzhou 510006, China}

\date{\today}

\begin{abstract}
We propose a scheme to realize controllable quantum state transfer and entanglement generation among transmon qubits  in the typical circuit QED setup based on adiabatic passage. Through designing the time-dependent driven pulses applied on the transmon qubits, we find that fast quantum sate transfer can be achieved between arbitrary two qubits and quantum entanglement among the qubits also can also be engineered. Furthermore, we numerically analyzed the influence of the decoherence on our scheme with the current experimental accessible systematical parameters. The result shows that our scheme is very robust against both the cavity decay and qubit relaxation, the fidelities of the state transfer and entanglement preparation process could be very high. In addition, our scheme is also shown to be insensitive to the inhomogeneous of qubit-resonator coupling
strengths.

\bigskip

\noindent Correspondence and requests for materials should be
addressed to Z.-Y.X. (email: zyxue@scnu.edu.cn)\\ or P.X. (email: xupengqh201461@163.cn) or F.M. (email: tianfengmei@gmail.com).
\end{abstract}

\pacs{03.67.Lx, 03.67.Bg, 85.25.Cp}

\maketitle

In the past two decades, many advances in quantum computation have been achieved on various   kinds of quantum systems. One of the crucial perquisites for realizing quantum information processing is quantum entanglement. Due to its non-locality and
non-classical correlations, quantum entanglement has been exploited in many applications \cite{nc}. To realize large scale quantum networks, one further needs quantum state transfer (QST) among different quantum nodes, which is the basic building blocks \cite{R10,R11,yang,jzhou,chen}. It is well-known that quantum information processing can be realized through the adiabatic evolution of the ground state of an initial Hamiltonian to that of a target Hamiltonian, i.e., the adiabatic passage \cite{R26,R27,R28}. It has been demonstrated that such technique is robust to the fluctuation of parameters as well as the decoherence due to spontaneous emission. Till now, QST and entanglement preparation have been widely explored in literature both theoretically \cite{R12,R13,R14,R15,R16,R17,R18,R19,R20, R21, R22,Sup20051,Sup20052,lyx1,lyx2,xuesr} and experimentally \cite{R23,R24,R25} based on such technique.

The recent rapid developments in circuit  QED has enabled this system as one of the
leading platforms for studying quantum computation and quantum simulation \cite{R291,Buluta,R292,R301,R302}. This system can also be easily scaled up to large scale and possesses long coherent time \cite{R31,R32}. One of the promising qubits in this context is the superconducting transmon qubit \cite{R33} which is immune to $1/f$ charge noise. The transmon qubit is a quantum LC oscillator with the inductor substituted by the Josephson junction and is designed to
suppress the charge noise to negligible values. The nonlinearity of the Josephson inductance further allows the oscillator to be weakly anharmonic, which make this system also be suitable for studying multi-level quantum operations. Recently, full quantum state tomography of a transmon as a three-level qutrit has been demonstrated \cite{R34}. The superconducting qubit control has also been realized with a combination of resonant microwave drives for achieving single-qubit rotations on nanosecond timescales. Furthermore, the transmon qubits connected to a transmission line resonator also provide a natural platform to study quantum optics and quantum computation. Many important experimental advances have been archived in this regard, including observation of Jaynes-Cummings ladder \cite{R35}, quantum trajectories \cite{R36} and Schr\"{o}dinger cat states \cite{R37}, and demonstration of quantum algorithms \cite{R38}, quantum teleportation \cite{R39}, geometric phase gates \cite{43Geometric}, Toffoli gate \cite{R41}, multi-qubit entanglement \cite{R42} and quantum error correction \cite{R40}. Besides, recent experiments \cite{R40,R44,R45} have also demonstrated that single- and two-qubit gates with fidelities can approach the fault-tolerant threshold for surface code, and thus provide the essential ingredients for realizing large-scale fault-tolerant quantum computation.

In this paper, we propose a scheme based on adiabatic passage to realize QST and quantum entanglement generation among three transmon qubits fabricated in a transmission line resonator. This method could also be simply generalized to many qubits case to achieve long-distant QST and multipartitie quantum entanglement. In our scheme, QST can be performed between arbitrary two qubits through applying approximate driven pulses. Moreover, based on tuning the time delay of the driven pulses, the transferred quantum state can be stabilized in a long time range, which is very helpful for further quantum information extraction with quantum non-demolition measurement. Furthermore, the degree of the final generated entanglement among the three transmon qubits can be tuned by changing the parameters of the driven pulse. In particular, we show the case for generating three qubit W state, which has many applications in quantum information processing. Finally, we numerically analyze the influence of the decoherence on our scheme, including the cavity decay and qubit relaxation. The result shows that our scheme is robust to their influence and the QST and entanglement generation could be achieved with very high fidelity.  Moreover, our method is also shown to be insensitive to the inhomogeneous qubit-resonator coupling strength.\\

\begin{figure}
\includegraphics[width=8cm]{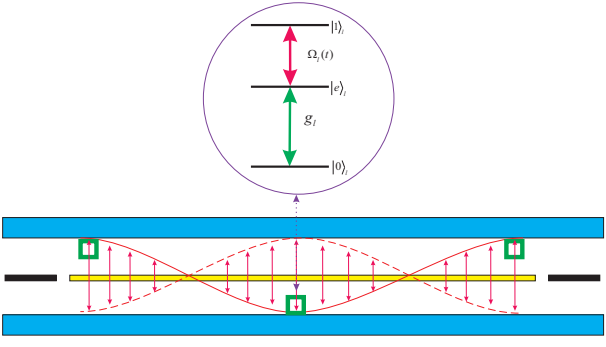}
\caption{{\bf Schematic of the superconducting system
consisting of three transmon qubits in a
transmission line resonator.} The ladder-type energy  level
configuration for the $l$th transmon qubit consists of one
auxiliary state $\left|e\right\rangle _{l}$ and the computational
basis states, $\left|0\right\rangle _{l}$ and
$\left|1\right\rangle _{l}$. The transition $\left|0\right\rangle
_{l}\leftrightarrow\left|e\right\rangle _{l}$ is resonantly
coupled to cavity field with a coupling constant $g_{l}$ and the
transition $\left|e\right\rangle
_{l}\leftrightarrow\left|1\right\rangle _{l}$ is driven by a
time-dependent microwave pulse with Rabi frequency
$\Omega_{l}(t)$.}\label{F1}
\end{figure}

\bigskip

{\bf \large Results}

{\bf The Building block.}
We consider three identical transmon qubits placed in a transmission line  resonator, and the resonator has single relevant mode with the frequency $w_{c}$ involving the qubit-photon interaction, as shown in Fig. \ref{F1}. We label the first three energy levels as the qubit state $\left|0\right\rangle $, an auxiliary state $\left|e\right\rangle $, and  the qubit state $\left|1\right\rangle $, respectively.  The transition $\left|e\right\rangle_{l}\leftrightarrow\left|1\right\rangle _{l}(l=1,2,3)$ is driven by $l$th time-dependent Rabi frequency and the transition
$\left|0\right\rangle _{l}\leftrightarrow\left|e\right\rangle_{l}$ is coupled to the cavity mode, while the $\left|0\right\rangle_{l}\leftrightarrow\left|1\right\rangle _{l}$ transition is
forbidden.
When $|w_{1e}-w_{0e}|$ is big enough, the driving field applied to the transition between $\left|1\right\rangle $ and $\left|e\right\rangle $ has small effect on the transition from $\left|0\right\rangle $ to $\left|e\right\rangle $ \cite{43Geometric}. The Hamiltonian of a single transmon qubit is given by
$H_{s}=4E_{C}(\hat{n}-n_{g})^{2}-E_{J}\cos\hat{\varphi}$,
where $\hat{n}$ and $\hat{\varphi}$ denote the number of Cooper pairs transferred between the islands and gauge-invariant phase difference between the superconductors. The effective offset charge $n_{g}=Q_{r}/2e+C_{g}V_{g}/2e$ is controlled by a gate electrode capacitively coupled to the island, where $C_{g}$ and $V_{g}$ represent the gate capacitance and voltage, $Q_{r}$ denotes the  environment-induced offset charge. The difference between the transmon and the Cooper pair box (CPB) is a shunting connection of the two superconductors via a large capacitance $C_{B}$. Via the additional capacitance $C_{B}$, the charging energy $E_C=e^{2}/(2C_{\varSigma})$ can be made small compared to the Josephson energy, where $C_{\varSigma}=C_{g}+2C_{J}+C_{B}$, $C_{J}$ is the capacitance of the Josephson tunnel junction. The Josephson energy  $E_{J}=2\tilde{E_{J}}\cos(\pi\Phi_{ext}/\Phi_{0})$ is tuned by means of an external magnetic flux $\Phi_{ext}$, with $\tilde{E_{J}}$ being the Josephson energy of a single junction. Compared with the CPB, the transmon is operated in the parameter regime $E_{J}\gg E_{C}$. The Hamiltonian of the superconducting transmission line resonator is
$H_{c}=\hbar w_{c}\left(a^{+}a+\frac{1}{2}\right)$,
where $w_{c}=1/\sqrt{L_{c}C_{c}}$ denotes the resonator frequency, and $a$ $(a^{+})$ represents the annihilation (creation) operator of the transmission line resonator. Under the rotating-wave approximation, the effective interaction Hamiltonian of the whole system can be written as
\begin{equation}
H_{I}=\sum_{l=1}^{3}[g_{l}a\left|e\right\rangle _{l}\left\langle 0\right|+\Omega_{l}(t)\left|1\right\rangle _{l}\left\langle e\right|]+h.c.,
\end{equation}
where we have assumed that $\hbar=1$, $g_{l}$ is the coupling constant between the cavity and the transmon, $\Omega_{l}(t)$ is the Rabi frequency for the transition $\left|e\right\rangle _{l}\leftrightarrow\left|1\right\rangle _{l}$ of the qubit $l$. Without loss of generality, we assume that the transmon qubit is driven by a time-dependent microwave pulse with Gaussian envelopes \cite{R461,R462}
\begin{equation}
\Omega_{l}(t)=\Omega_{0l}e^{-(t-\tau_{l})^{2}/T_{l}^{2}},
\end{equation}
where $\Omega_{0l}$, $\tau_{l}$ and $T_{l}$ are pulse amplitude, time delay and operation duration.  In the following, the parameters and the shape of the driven Gaussian pulses will be engineered for achieving certain target quantum information processing, which is within the current state of the art circuit QED technology. Recently, based on similar engineering on the external driven Gaussian microwave pulses, non-adiabatic \cite{Nonad20021,Nonad20022,Nonad20023} non-abelian geometric phase has been observed with transmon qubit \cite{43Geometric,xuepra}.

\bigskip

{\bf Quantum state transfer.}
We now show how to engineer the driven pulses based on adiabatic passage \cite{R26} to realize QST among three transmon qubits. In particular, we separately discuss two situations with the qubit-resonator coupling strengths are homogeneous and inhomogeneous. The result shows that our scheme is robust to the inhomogeneous  of qubit-resonator coupling strength. The total excitation operator in our system can be written as
$N_{e}=\sum_{l=1}^{3}(\left|e\right\rangle _{l}\left\langle e\right|+\left|1\right\rangle _{l}\left\langle 1\right|)+a^{+}a$,
where $N_{e}$ commutes with $H_{I}$ so that the excitation number is conserved during the dynamical evolution. Here, we assume a single excitation is coherently exchanged between the qubit and resonator. The resonator can be the quantum bus that mediates interactions between the qubits, which can realize the QST among different qubits. The subspace in our scheme is the single excitation subspace, i.e., $N_{e}=1$.

Assume the initial state of the system is $\left|0\right\rangle _{c}\left|1\right\rangle _{1}\left|0\right\rangle _{2}\left|0\right\rangle _{3}$, where the transmon qubit 1 is prepared in the state $\left|1\right\rangle $, transmon qubits 2 and 3 are in the state
$\left|0\right\rangle $, and the cavity field is in the vacuum state. In this situation, the system evolves within this subspace is spanned by seven basis
\begin{eqnarray}
&& \left|\phi_{1}\right\rangle =\left|0\right\rangle _{c}\left|1\right\rangle _{1}\left|0\right\rangle _{2}\left|0\right\rangle _{3},\quad
\left|\phi_{2}\right\rangle =\left|0\right\rangle _{c}\left|0\right\rangle _{1}\left|1\right\rangle _{2}\left|0\right\rangle _{3},\nonumber \\
&& \left|\phi_{3}\right\rangle =\left|0\right\rangle _{c}\left|0\right\rangle _{1}\left|0\right\rangle _{2}\left|1\right\rangle _{3},\quad
\left|\phi_{4}\right\rangle =\left|1\right\rangle _{c}\left|0\right\rangle _{1}\left|0\right\rangle _{2}\left|0\right\rangle _{3},\nonumber \\
&& \left|\phi_{5}\right\rangle =\left|0\right\rangle _{c}\left|e\right\rangle _{1}\left|0\right\rangle _{2}\left|0\right\rangle _{3},\quad
\left|\phi_{6}\right\rangle =\left|0\right\rangle _{c}\left|0\right\rangle _{1}\left|e\right\rangle _{2}\left|0\right\rangle _{3},\nonumber \\
&& \left|\phi_{7}\right\rangle =\left|0\right\rangle _{c}\left|0\right\rangle _{1}\left|0\right\rangle _{2}\left|e\right\rangle _{3},
\end{eqnarray}
where $\left|\phi_{4}\right\rangle \sim \left|\phi_{7}\right\rangle $ are the intermediate states.  It is easy to verify that the following two states
are the eigenstates of the effective Hamiltonian with zero eigenvalue
\begin{eqnarray}
\left|D(0)\right\rangle _{3}&=&\left|0\right\rangle _{c}\left|0\right\rangle _{1}\left|0\right\rangle _{2}\left|0\right\rangle _{3}\\
\left|D(t)\right\rangle _{3}&=&N_{3}[g_{1}\Omega_{2}(t)\Omega_{3}(t)\left|\phi_{1}\right\rangle + g_{2}\Omega_{1}(t)\Omega_{3}(t)\left|\phi_{2}\right\rangle\nonumber \\
&& +g_{3}\Omega_{1}(t)\Omega_{2}(t)\left|\phi_{3}\right\rangle - \Omega_{1}(t)\Omega_{2}(t)\Omega_{3}(t)\left|\phi_{4}\right\rangle],\nonumber
\end{eqnarray}
where $N_{3}$ is the normalization factor. The states $\left|D(0)\right\rangle _{3}$ and $\left|D(t)\right\rangle _{3}$ are dark states since  they have no dynamics under the Hamiltonian, and thus the excited level $\left|e\right\rangle _{l} (l=1,2,3)$ is unpopulated. When the system is in the state $\left|0\right\rangle _{c}\left|0\right\rangle _{1}\left|0\right\rangle _{2}\left|0\right\rangle _{3}$, the dynamical evolution of the system will be frozen.

From the above dark state formalism, it is easy to find that, if one can simultaneously engineer the three driven Gaussian pulses so that initially $\{\Omega_{3}(t),\Omega_{2}(t)\} \gg\Omega_{1}(t)$ and finally $\{\Omega_{3}(t),\Omega_{1}(t)\} \gg\Omega_{2}(t)$, then the population transfer could be achieved from $\left|\phi_{1}\right\rangle$ to $\left|\phi_{2}\right\rangle$. The detailed evolution path and the driven pulses needed in the above transfer can be described as the following. First,
\begin{equation}
 \left|\phi_{1}\right\rangle\stackrel{\Omega_{1}}
 {\longrightarrow}\left|\phi_{5}\right\rangle\stackrel{g_{1}}
 {\longrightarrow}\left|\phi_{4}\right\rangle.
\end{equation}
Then, the photon is further absorbed by the second qubit, the system will further evolve according to
\begin{equation}
\left|\phi_{4}\right\rangle\stackrel{g_{2}}
{\longrightarrow}\left|\phi_{6}\right\rangle\stackrel{\Omega_{2}}
{\longrightarrow}\left|\phi_{2}\right\rangle.
\end{equation}
Based on this observation, one can realize QST between two transmon qubits. Suppose that the original quantum information is encoded in an arbitrary unknown quantum state $\left|\phi\right\rangle =a\left|0\right\rangle +b\left|1\right\rangle $ carried by the transmon qubit 1, where $a$ and $b$ are the normalized coefficients. As shown above, the zero excitation state $\left|0\right\rangle _{c}\left|0\right\rangle _{1}\left|0\right\rangle _{2}\left|0\right\rangle _{3}$ is frozen, the one-excitation state can be swapped between the transmon qubit 1 and 2, then the coherent quantum state $\left|\phi\right\rangle =a\left|0\right\rangle +b\left|1\right\rangle $ could be finally transferred from transmon qubit 1 to 2.

Similarly, one can realize the QST from transmon qubit 2 to 3. In this case, based on observing the dark state formalism, when the three driven Gaussian pulses are engineered simultaneously so that initially $\{\Omega_{3}(t),\Omega_{1}(t)\} \gg\Omega_{2}(t)$ and finally $\{\Omega_{1}(t),\Omega_{2}(t)\} \gg\Omega_{3}(t)$, the population from the initial state $\left|\phi_{2}\right\rangle $ would be transferred to the target state $\left|\phi_{3}\right\rangle $. The detailed evolution process can be written as first
\begin{equation}
 \left|\phi_{2}\right\rangle\stackrel{\Omega_{2}}
 {\longrightarrow}\left|\phi_{6}\right\rangle\stackrel{g_{2}}
 {\longrightarrow}\left|\phi_{4}\right\rangle,
\end{equation}
and then the photon is further absorbed by the third qubit
\begin{equation}
 \left|\phi_{4}\right\rangle\stackrel{g_{3}}
 {\longrightarrow}\left|\phi_{7}\right\rangle\stackrel{\Omega_{3}}
 {\longrightarrow}\left|\phi_{3}\right\rangle,
\end{equation}
which is the coherent QST between the qubit 2 and 3. In the whole process, one can find that the cavity state and the qubit state $|e\rangle$ are the intermediate states. It is worth to point out that, our method is also can be generalized to realize QST between arbitrary two qubits, including one particular qubit to the one that is far away from this qubit, which is of great significance to the large scale quantum computation in the future.

\begin{figure}
\includegraphics[width=\columnwidth]{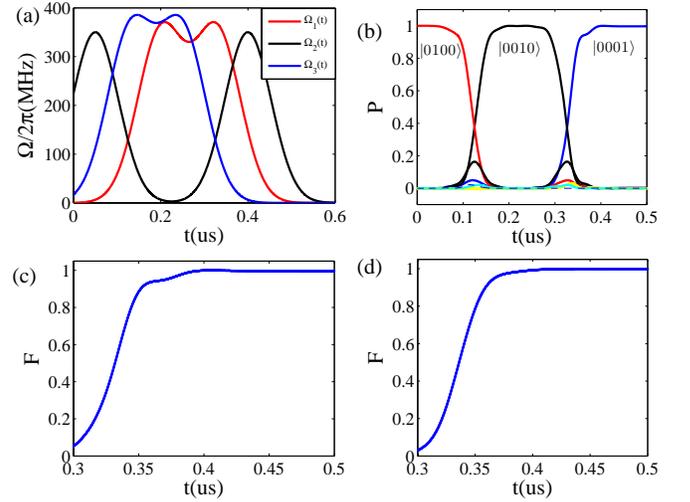}
\caption{{\bf Numerical results for quantum state transfer from qubit 1 to 2 and then to 3.} (a) The shape of the driven pulses with $\Omega_{ij,i=1,2,3;j=1,2}/2\pi= 350$ MHz, $\tau_{1}=0.35$ us, $\tau_{2}=0.58$ us, $\tau_{3}=0.2$ us, $\tau_{4}=0.65$ us, $\tau_{5}=0.28$ us, $\tau_{6}=0.5$ us, $T_{i=1,2,...,6}=75$ ns.
(b) Time evolution of the populations in different states during the
population transfer from $\left|0100\right\rangle $ to $\left|0001\right\rangle $,
with $g_{l}/{2\pi}=200$ MHz. (c) Fidelity F versus the time $t$, with $\Delta{g}/{2\pi}=0$. (d) Fidelity F versus the time $t$, with the deviation $\Delta{g}/{2\pi}=40$ MHz ($g_{1}/{2\pi}=180$ MHz, $g_{2}/{2\pi}=200$ MHz, $g_{3}/{2\pi}=160$ MHz).}\label{F2-(a)-(b)-(c)-(d)}
\end{figure}

In the following, we will show the method to design the driven Gaussian pulse sequence with their parameters satisfying the above requirements. For this purpose, we firstly assume the forms of the time-dependent driven Gaussian pulses are chosen as
\begin{eqnarray}
\Omega_{1}(t) & = & \Omega_{11}e^{-(t-\tau_{1})^{2}/T_{1}^{2}}+\Omega_{12}e^{-(t-\tau_{2})^{2}/T_{2}^{2}},\nonumber\\
\Omega_{2}(t)  & = & \Omega_{21}e^{-(t-\tau_{3})^{2}/T_{3}^{2}}+\Omega_{22}e^{-(t-\tau_{4})^{2}/T_{4}^{2}},\\
\Omega_{3}(t) & = & \Omega_{31}e^{-(t-\tau_{5})^{2}/T_{5}^{2}}+\Omega_{32}e^{-(t-\tau_{6})^{2}/T_{6}^{2}}.\nonumber
\end{eqnarray}
Through substituting the above forms into the systemic Hamiltonian, we numerically calculate the systemic evolution and go to find the optimal parameters where the fidelity of our scheme is maximal. In Fig. \ref{F2-(a)-(b)-(c)-(d)}(a),  the three optimal Gaussian pulses $\Omega_{1,2,3}(t)$ for achieving QST from qubit 1 to 2 and then to 3 at a fixed time delay are plotted. In Fig. \ref{F2-(a)-(b)-(c)-(d)}(b), the detailed population transfer of the QST process is calculated. One can find that the QST between two qubits can be obtained within a time duration $t_{f}=150$ ns, which is faster than  that of the atomic system \cite{Riebe}. However, the transfer time can be much shorter through optimizing the driven pulses and improving the coupling strength between the qubit and resonator. We also numerically find the optimal parameter to make the QST much faster at a cost of reducing the fidelity. The good news is that the fidelity can be still higher than 0.9. Besides, during the transfer process, the intermediate states emerge in a very short time range compared to the decoherence time, which is the reason why our scheme has high fidelity even in the presence of decoherence. Another interesting characteristic during the system evolution is that the residence time on the populations of the quantum states is controllable through engineering the driven pulses, which provide an extra freedom to manipulate the QST in such framework.

In Fig. \ref{F2-(a)-(b)-(c)-(d)}(c), we have plotted the fidelity of the
above QST from transmon qubit 1 to 3 through 2. The fidelity is formulated as $F(t)=\left(Tr\sqrt{\rho_{f}^{1/2}\rho(t)\rho_{f}^{1/2}}\right)^{2}$, where $\rho(t)$ and $\rho_{f}$ are the density matrixes of the evolved state at the end of the pulse operation and the ideal final state respectively. One can find that the fidelity can almost approach one. Actually, the fidelity for QST from qubits 1 to 3 is much higher. Note that the influence of the inhomogeneous qubit-resonator coupling strengths on our scheme caused by the imperfection in the practical fabrication is a very important issue needed to be addressed. With a typical choice on the inhomogeneous qubit-resonator coupling strengths, we numerically calculate the corresponding fidelity and analyze the performance in this case in Fig. \ref{F2-(a)-(b)-(c)-(d)}(d). The result shows that the fidelity of our scheme also can approach one, and thus very robust.

\begin{figure}
\includegraphics[width=\columnwidth]{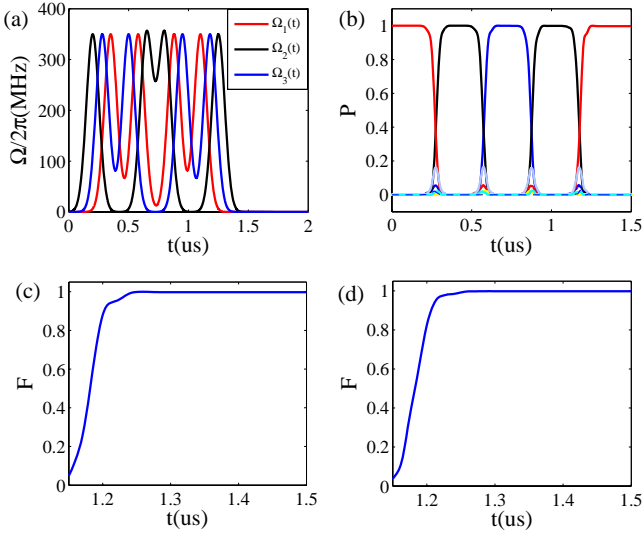}
\caption{{\bf Numerical results for quantum state transfer from  qubit 1 to 3 through 2 and and then back to 1.} (a) The shape of the driven pulses. (b) Time evolution of the corresponding populations for the different quantum states, with $g_{l}/{2\pi}=200$ MHz. (c) The fidelity F versus time $t$, with $\Delta{g}/2\pi=0$. (d) Fidelity F versus the time $t$, with the deviation $\Delta{g}/{2\pi}=40$ MHz ($g_{1}/{2\pi}=180$ MHz, $g_{2}/{2\pi}=200$ MHz, $g_{3}/{2\pi}=160$ MHz).}\label{F3-(a)-(b)-(c)-(d)}
\end{figure}

In Fig. \ref{F3-(a)-(b)-(c)-(d)}, we further numerically demonstrates that the present protocol can also be generalized to realize QST from transmon qubit 1 to 3 through 2 and and then back to 1, including numerically finding the optimal Gaussian pulses, calculating the population transfer process and the fidelities for homogeneous and inhomogeneous qubit-resonator coupling strengths. The result shows that our scheme also can work very well even in this case. The final achieved fidelity could nearly approach one, even in the presence of an inhomogeneous case. Therefore, our scheme for QST using driven pulses is very robust, including working well with resonant and non-resonant, homogeneous and inhomogeneous qubit-resonator coupling.

\bigskip

{\bf Entangled state generation.}
We next consider using driven pulses to robustly generate quantum entanglement among the three transmon qubits placed in the transmission line resonator. There have been some entanglement generation schemes in similar circuit QED setups. Firstly, Tsomokos \cite{Tsomokos} has presented a scheme of entanglement generation that N charge (flux) qubits are coupled capacitively (inductively). However, due to the fact that the coupling between every two qubits is required and each qubit is operated at its degeneracy point, it needs more time to achieve the entanglement and more fragile under the decoherence than our scheme. Secondly, Galiautdinov and Martinis \cite{Galiautdinov} have introduced another scheme that the maximal entanglement is generated in tripartite system with pairwise coupling $g(XX+YY)+\widetilde{g}ZZ$. It is realized in Josephson phase qubits with capacitive and inductive couplings. However, the decoherence time of the phase qubit is shorter than transmon qubit in our scheme and the interaction can not  be realized similarly using transmon qubits.  We assume that the initial state of the system is $|\phi_{1}\rangle$.  When the driven pulses and the qubit-resonator couplings are applied, the evolution process is described as
\begin{eqnarray}
|\phi_{1}\rangle\stackrel{\Omega_{1}}{\longrightarrow}|\phi_{5}\rangle\stackrel{g_{1}}
{\longrightarrow}|\phi_{4}\rangle
\begin{cases}
\stackrel{g_{1}}{\longrightarrow}|\phi_{5}\rangle\stackrel{\Omega_{1}}{\longrightarrow}|\phi_{1}\rangle,\\
\stackrel{g_{2}}{\longrightarrow}|\phi_{6}\rangle\stackrel{\Omega_{2}}{\longrightarrow}|\phi_{2}\rangle, \\
\stackrel{g_{3}}{\longrightarrow}|\phi_{7}\rangle\stackrel{\Omega_{3}}{\longrightarrow}|\phi_{3}\rangle.
\end{cases}
\end{eqnarray}
Firstly, the driven pulse with Rabi frequency $\Omega_{1}(t)$ drives the system from $|\phi_{1}\rangle$ into the state $|\phi_{5}\rangle$, then evolves into the state $|\phi_{4}\rangle$ through the coupling between transmon qubit 1 and the resonator. Note that the state $|\phi_{4}\rangle$ is just one-photon state. So, from now on, all the couplings between the three qubits and the resonator will dominant over the evolution, make $|\phi_{4}\rangle$ evolve into $|\phi_{5,6,7}\rangle$ with equal weights. After that, three driven pulses with Rabi frequencies $\Omega_{1,2,3}(t)$ would bring $|\phi_{5,6,7}\rangle$ into $|\phi_{1,2,3}\rangle$ with equal weights, leaving the cavity in the vacuum state. Then we get the entangled state of the three transmon qubits as
\begin{equation}
|W\rangle =\frac{1}{\sqrt{3}}(|100\rangle_{123}+|010\rangle_{123}+|001\rangle_{123}),
\end{equation}
which is a W state and can be employed to complete various quantum information processing tasks.

\begin{figure}
\includegraphics[width=\columnwidth]{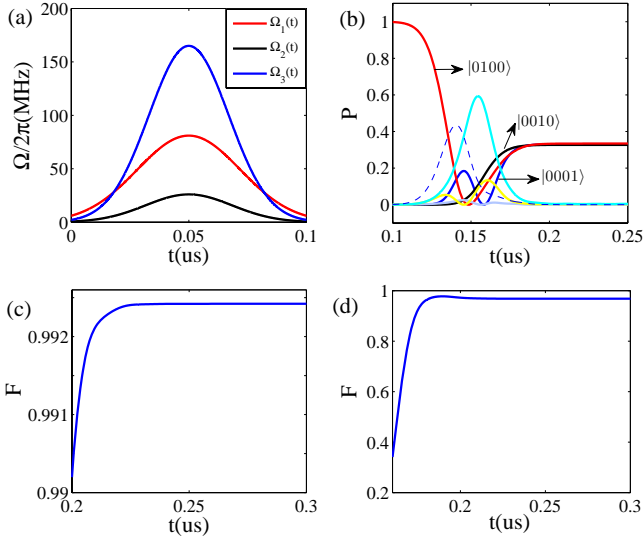}
\caption{{\bf Numerical results for entangled state generation.} (a) The shape of  of driven pulses with $\Omega^{'}_{11}/{2\pi}=81$ MHz, $\Omega^{'}_{21}/{2\pi}=26$ MHz, $\Omega^{'}_{31}/{2\pi}=165$ MHz,
$\tau_{1}=\tau_{2}=\tau_{3}=0.15$ us, $T_{1}=31$ ns, $T_{2}=26$ ns,
$T_{3}=24$ ns. (b) Time evolution of the populations for the different quantum
states, with $g_{l}/{2\pi}=200$ MHz.
(c) The fidelity F versus time $t$, with $\Delta{g}/2\pi=0$. (d) Fidelity F versus the time $t$, with the deviation $\Delta{g}/{2\pi}=40$ MHz ($g_{1}/{2\pi}=180$ MHz, $g_{2}/{2\pi}=200$ MHz, $g_{3}/{2\pi}=160$ MHz).}\label{F4-(a)-(b)-(c)-(d)}
\end{figure}

The detailed performance of the above quantum entanglement generation is further analyzed through numerically designing the driven pulses. For this purpose, the form of three optimal time-dependent driven Rabi frequencies are chosen as
\begin{eqnarray}
&& \Omega^{'}_{1}(t)  = \Omega^{'}_{11}e^{-(t-\tau_{1})^{2}/T_{1}^{2}},\nonumber\\
&& \Omega^{'}_{2}(t)  = \Omega^{'}_{21}e^{-(t-\tau_{2})^{2}/T_{2}^{2}},\nonumber\\
&& \Omega^{'}_{3}(t)  = \Omega^{'}_{31}e^{-(t-\tau_{3})^{2}/T_{3}^{2}}.
\end{eqnarray}
In Fig. \ref{F4-(a)-(b)-(c)-(d)}(a-b), we have plotted the optimal three driven pulses and the time evolution of the systemic populations. In contrast, in Fig. \ref{F4-(a)-(b)-(c)-(d)}(a-b), one can find that there are three states left finally and their coherent superposition leads to an W state. The results also show that the interaction time required for generating such entanglement among the three transmon qubits is about $85$ ns, which is very fast compared to previous schemes for quantum entanglement preparation. In Fig. \ref{F4-(a)-(b)-(c)-(d)}(c), we further plot the fidelity of the evolved states as function of the time and find that the fidelity of the final entanglement could be higher than 0.99. As shown in \ref{F4-(a)-(b)-(c)-(d)}(d), this conclusion holds even for the nonidentical qubit-resonator coupling strengthes.

Furthermore, this method can also be directly employed to generate N-qubit W state. For instance, N-transmon qubits are fabricated in a transmission line resonator. All the transmon qubits are resonate with the single-mode resonator and driven by the time-dependent pulses. The effective Hamiltonian for the system is
\begin{equation}
H_{I}=\sum_{l=1}^{N}[g_{l}a\left|e\right\rangle _{l}\left\langle 0\right|+\Omega_{l}(t)\left|1\right\rangle _{l}\left\langle e\right|]+h.c..
\end{equation}
We verify that the following two states
\begin{eqnarray}
|D(0)\rangle_{N}&=&|0\rangle_{1}|0\rangle_{2}\cdot\cdot\cdot|0\rangle_{N}|0\rangle_{c},\notag\\
|D(t)\rangle_{N}&=&N_{n}\left[\frac{g_{1}}{\Omega_{1}(t)}|1\rangle_{1}
\prod_{l=2}^{n}|0\rangle_{l}|0\rangle_{c}-\prod_{l=1}^{n}|0\rangle_{l}|1\rangle_{c}\right.\notag\\
&&\left.+\sum_{l=2}^{n}\frac{g_{l}}{\Omega_{l}(t)}|0\rangle_{1}\cdot\cdot\cdot
|0\rangle_{l-1}|1\rangle_{c}\cdot\cdot\cdot|0\rangle_{n}\right]
\end{eqnarray}
are eigenstates of the Hamiltonian with zero eigenvalue. Here, $N_{n}$ is a normalization factor. The states $|D(0)\rangle_{N}$ and $|D(t)\rangle_{N}$ are dark states since the excited level $|D(0)\rangle_{l}$ is unpopulated. Through optimizing the time-dependent driven pulses applied on the transmon qubits, we can achieve the fast quantum state transfer and quantum entanglement among different qubits. Actually, one also can find that the degree of the above entanglement also can be engineered to a particular value through designed approximate driven pulses. Moreover, different from the previous methods, the Bell state and W state could be prepared between arbitrary two and three qubits with such method. Such feature is very helpful for achieving large scale quantum computation in a quantum network.

\bigskip

{\bf Discussion.}
At this stage, we take into account the cavity decay and qubit
relaxation and analyze their influences on  the quantum state
transfer and quantum entanglement generation. For this purpose,
the master equation is employed to described the  above
decoherence process, which can be written as
\begin{eqnarray}
\frac{d\rho}{dt}&=&-i[H_{I},\rho]+{\kappa \over 2} \mathcal{L}[a]+{1 \over 2} \sum_{l=1}^{3}\{\gamma_{l,1e}\mathcal{L}[\sigma_{l,1e}^-]\notag\\
&& +\gamma_{l,e0}\mathcal{L} [\sigma_{l, e0}^-] +\varGamma_{l,1}\mathcal{}L[\sigma_{l,1}]
+\varGamma_{l,e}\mathcal{L}[\sigma_{l,e}]\},
\end{eqnarray}
where $\kappa$ is the decay rate of the cavity, $\varGamma_{l,1}$ $\left(\varGamma_{l,e}\right)$
is the dephasing rate of the qubit $l$ with the level $\left|1\right\rangle $
$\left(\left|e\right\rangle \right)$,  $\gamma_{l,1e}$ and $\gamma_{l,e0}$ are the energy relaxation rates
for the qubit $l$ with the decay path $\left|1\right\rangle \rightarrow\left|e\right\rangle$ and $\left|e\right\rangle \rightarrow\left|0\right\rangle$, respectively; $\mathcal{L}[A]=2A\rho A^{+}-A^{+}A\rho-\rho A^{+}A$, $\sigma_{l,ij}^{-}=\left|j\right\rangle _{l}\left\langle i\right|$ $\sigma_{l,ij}^{+}=\left|i\right\rangle _{l}\left\langle j\right|$,  and $\sigma_{l,k}=\left|k\right\rangle _{l}\left\langle k\right|$  $(k=1,e)$.

\begin{figure}
\includegraphics[width=\columnwidth]{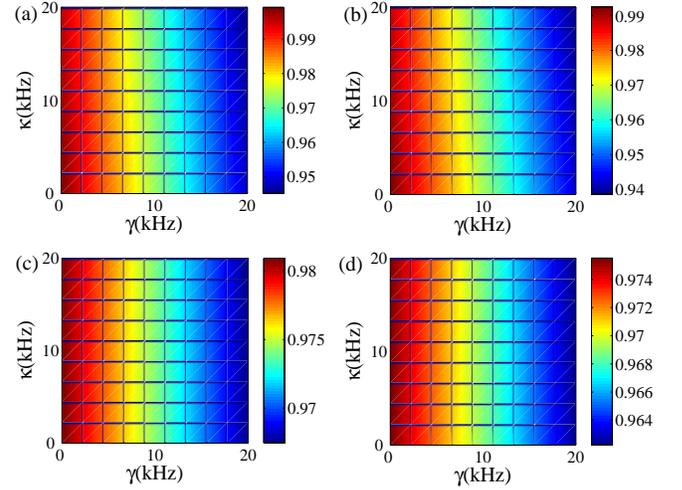}
\caption{{\bf Numerical simulation of the influence of the decohenrence.} Fidelities versus $\gamma$ and
$\kappa$ at the end of the operation time with respect to the target state based on the master equation (13) for the quantum
state transfer (a,b) and the three-qubit  entanglement generation
(c,d) in the homogeneous and inhomogeneous qubit-resonator
couplings.}\label{F5-(a)-(b)-(c)-(d)}
\end{figure}

For simplicity, we assume $\varGamma_{1}=\gamma/2$, $\varGamma_{e}=\gamma/4$,
$\gamma_{1e}=\gamma$ and $\gamma_{e0}=\gamma/2$ \cite{R47}. Based on numerically solving the master equation, in Fig. \ref{F5-(a)-(b)-(c)-(d)}, we calculate the fidelity of the population transfer and
the generation of entanglement among the three qubits in the presence of the decoherence. The results show that the fidelity decreases with the increase of the cavity decay and qubit relaxation rate.  As both the life time of the photons in the resonator and the coherence time of the transmon qubit can approach $20$ us  \cite{R47}, it is easy to check that the parameters involved in our numerical calculation is within the experimental accessible parameter regime. With a typical choice of $g_{l}/{2\pi}=200$ MHz, $\kappa/{2\pi}=20$ kHz, $\gamma_{1e}/{2\pi}=20$ kHz, $\varGamma_{1}/{2\pi}=10$ kHz, $\gamma_{e0}/{2\pi}=10$ kHz, $\varGamma_{e}/{2\pi}=5$ kHz, the fidelity of the final state can be higher than 0.94. The decoherence has a smaller effect on entanglement generation than on state transfer. The reason is that the quantum state transfer need much longer time than the entanglement generation, this is due to the fact that we need to insert delay pulses to further modulate the quantum dynamics for fulfilling the whole transfer process, while the entanglement generation process is straightforward.


In summary, based on engineering external driven microwave pulses and adiabatical passage, we have proposed a controllable method to realize high-fidelity QST and quantum entanglement among three superconducting qubits embedded in a transmission line resonator, which can be achieved with fast speed and high fidelity even in the presence of decoherence. Moreover, we also demonstrate that our scheme is also very robust to the inhomogeneousness of qubit-resonator coupling strength. In addition, our method can be readily scaled up to realize long-distance QST and multipartite quantum entangled generation. Finally, our proposal can also be applied to other types of superconducting qubits. Therefore, it is expected that our scheme could add a robust means for circuit QED to realize large-scale quantum computation and quantum simulation.

\bigskip

\noindent {\bf Acknowledgements} 
We acknowledge helpful discussions with Y. Yu
and S. L. Zhu. This work was supported by the SKPBR of China
(Grant No. 2011CB922104), the NFRPC (No. 2013CB921804),
 the NSFC (Grants No. 11125417, No. 11474153, and No. 11274156), and the PCSIRT (Grant No. IRT1243).

\end{document}